\documentclass[aps,preprint,showpacs,groupedaddress,floatfix]{revtex4}
\usepackage{graphicx}
\usepackage{dcolumn}
\usepackage{bm}
\begin{document}
\title{Relativistic Random-Phase Approximation with
density-dependent meson-nucleon couplings}
\author{T. Nik\v si\' c}
\affiliation{Physik-Department der Technischen Universit\"at M\"unchen,
D-85748 Garching, Germany}
\affiliation{ Physics Department, University of
Zagreb, 10000 Zagreb, Croatia}

\author{D. Vretenar}
\affiliation{ Physics Department, University of
Zagreb, 10000 Zagreb, Croatia}

\author{P. Ring}
\affiliation{Physik-Department der Technischen Universit\"at M\"unchen,
D-85748 Garching, Germany}
\vspace{1cm}
\date{\today}

\begin{abstract}
The matrix equations of the relativistic random-phase approximation
(RRPA) are derived for an effective Lagrangian characterized by 
density-dependent meson-nucleon vertex functions. The explicit
density dependence of the meson-nucleon couplings introduces 
rearrangement terms in the residual two-body interaction, 
that are essential for a quantitative description of excited
states. Illustrative calculations of the isoscalar monopole,
isovector dipole and isoscalar quadrupole response of $^{208}$Pb,
are performed in the fully self-consistent RRPA framework, based
on effective interactions with a phenomenological density 
dependence adjusted to nuclear matter and ground-state properties
of spherical nuclei. The comparison of the RRPA results on 
multipole giant resonances with experimental data constrains the 
parameters that characterize the isoscalar and isovector channel
of the density-dependent effective interactions.
\end{abstract}

\maketitle

\section{\label{secI}Introduction}
The success of models based on the Relativistic 
Mean Field (RMF)~\cite{Rin.96} 
approximation in describing structure phenomena, not only
in nuclei along the valley of $\beta$-stability, but also 
in exotic nuclei with extreme isospin
values and close to the particle drip lines, has also renewed the interest
in theoretical studies based on the 
Relativistic Random Phase Approximation (RRPA). Although several 
RRPA implementations have been available since the eighties, 
only very recently RRPA-based calculations
have reached a level on which a quantitative comparison with
experimental data became possible. Two points are
essential for the successful application of the RRPA in the
description of dynamical properties of finite nuclei: (i) the use
of effective Lagrangians with nonlinear self-interaction terms, 
and (ii) the fully consistent treatment of the Dirac sea
of negative energy states. Many studies over the last decade have
shown that the inclusion of nonlinear meson terms in meson-exchange
RMF models, or nonlinear nucleon self-interaction terms in relativistic
point-coupling models, is absolutely necessary in order to reproduce 
ground-state properties of spherical and deformed nuclei on a
quantitative level. Techniques which enable the inclusion
of nonlinear meson interaction terms in the RRPA framework, however, 
have been developed only recently in the calculation of the 
relativistic linear response~\cite{MGT.97}, and in the 
solution of the RRPA-matrix equation~\cite{VWR.00}.
For a quantitative description of excited states, 
the RRPA configuration space must include not only the usual particle-hole
states, but also pair-configurations formed from occupied states
in the Fermi sea and empty negative-energy states in the Dirac sea.
Even though it was known for a long time that the inclusion of
configurations built from occupied positive-energy states and
empty negative-energy states is essential for current conservation
and the decoupling of spurious states~\cite{DF.90}, 
only recently it has been
shown that the fully consistent inclusion of the Dirac sea of
negative energy states in the RRPA is essential for a quantitative
comparison with the experimental excitation energies of 
giant resonances~\cite{VWR.00,Rin.01}.

The RRPA with nonlinear meson interaction terms, and with a
configuration space that includes the Dirac sea of negative-energy 
state, has been very successfully employed in studies of 
nuclear compressional modes~\cite{VWR.00,Pie.01,MGW.01}, 
of multipole giant resonances and
of low-lying collective states in spherical nuclei~\cite{Mawa.01},
of the evolution of the low-lying isovector dipole response in nuclei 
with a large neutron excess~\cite{Vre.01,Vre.01b}, and of 
toroidal dipole resonances~\cite{Vre.02}.

An interesting alternative to the highly successful RMF models
with nonlinear self-interaction terms, is an effective 
hadron field theory with medium dependent
meson-nucleon vertices. Such an approach retains the basic
structure of the relativistic mean-field framework, 
but could be more directly related to the
underlying microscopic description of nuclear interactions.
In particular, the density dependent relativistic hadron field 
(DDRH) model~\cite{FLW.95} has been successfully applied
in the calculation of nuclear matter and ground-state properties of 
spherical nuclei~\cite{TW.99}, and extended to
hypernuclei~\cite{KHL.00}, neutron star matter~\cite{HKL.01a}, and
asymmetric nuclear matter and exotic nuclei~\cite{HKL.01}. Very 
recently, in Ref.~\cite{NVFR.02} we have extended the 
relativistic Hartree-Bogoliubov (RHB) model~\cite{PVL.97} to include
density dependent meson-nucleon couplings. The effective Lagrangian is
characterized by a phenomenological density dependence of the $\sigma$,
$\omega$ and $\rho$ meson-nucleon vertex functions, adjusted to properties
of nuclear matter and finite nuclei. It has been shown that, in comparison 
with standard RMF effective interactions with nonlinear meson-exchange
terms, the density-dependent meson-nucleon couplings significantly
improve the description of symmetric and asymmetric nuclear matter, 
and of isovector ground-state properties of $N\neq Z$ nuclei. This is,
of course, very important for the extension of RMF-based models 
to exotic nuclei far from $\beta$-stability (description of the 
neutron skin, the neutron halo, pygmy isovector dipole resonances), and
for applications in the field of nuclear astrophysics. 

In this work we derive the RRPA with density-dependent meson-nucleon 
couplings. Just as in the static case the single-nucleon Dirac 
equation includes the additional rearrangement self-energies that 
result from the variation of the vertex functionals with respect 
to the nucleon field operators, the explicit density dependence
of the meson-nucleon couplings introduces rearrangement terms
in the residual interaction of the RRPA.
The rearrangement contribution 
is essential for a quantitative analysis of 
excited states in the RRPA framework. In Sec.~\ref{secII} we present
the formalism of the relativistic RPA with density-dependent 
meson-nucleon couplings, and derive the RRPA equations in the small 
amplitude limit of the time-dependent RMF. The results of an 
illustrative calculation of multipole giant resonances in 
$^{208}$Pb are analyzed in Sec.~\ref{secIII}. Section~\ref{secIV}
contains the summary and the conclusions.


\section{\label{secII}Formalism of the Relativistic Random-Phase 
Approximation with density-dependent meson-nucleon couplings}
The standard density dependent relativistic hadron field (DDRH) 
model~\cite{FLW.95} for nuclear matter and finite nuclei
is defined by the Lagrangian density
\begin{eqnarray}
{\cal L} &=&\bar{\psi}\left( i{\bm \gamma} \cdot 
{\bm \partial} -m\right) \psi ~+~\frac{1%
}{2}(\partial \sigma )^{2}-\frac{1}{2}m_{\sigma }^{2}\sigma ^{2}  \nonumber \\
&&-~\frac{1}{4}\Omega _{\mu \nu }\Omega ^{\mu \nu }+\frac{1}{2}m_{\omega
}^{2}\omega ^{2}~-~\frac{1}{4}{\vec{{\rm R}}}_{\mu \nu }{\vec{{\rm R}}}^{\mu
\nu }+\frac{1}{2}m_{\rho }^{2}\vec{\rho}^{\,2}~-~\frac{1}{4}{\rm F}_{\mu \nu
}{\rm F}^{\mu \nu }  \nonumber \\
&&-~g_{\sigma }\bar{\psi}\sigma \psi ~-~g_{\omega }\bar{\psi}\gamma \cdot
\omega \psi ~-~g_{\rho }\bar{\psi}\gamma \cdot \vec{\rho}\vec{\tau}\psi ~-~e%
\bar{\psi}\gamma \cdot A\frac{(1-\tau _{3})}{2}\psi \;.
\label{lagrangian}
\end{eqnarray}
Vectors in isospin space are denoted by arrows, and bold-faced
symbols indicate vectors in ordinary three-dimensional space.
The Dirac spinor $\psi$ denotes
the nucleon with mass $m$.  $m_\sigma$, $m_\omega$, and
$m_\rho$ are the masses of the $\sigma$-meson, the
$\omega$-meson, and the $\rho$-meson.  $g_\sigma$,
$g_\omega$, and $g_\rho$ are the corresponding coupling
constants for the mesons to the nucleon. $e^2 /4 \pi =
1/137.036$. The coupling constants and the mass of the $\sigma$-meson
are treated as free parameters, adjusted to reproduce nuclear matter properties 
and ground-state properties of finite nuclei.
$\Omega ^{\mu \nu }$, $\vec{R}^{\mu \nu }$, and $F^{\mu \nu }$ are the field
tensors of the vector fields $\omega $, $\rho $, and of the photon:
\begin{eqnarray}
\Omega ^{\mu \nu } &=&\partial ^{\mu }\omega ^{\nu }-\partial ^{\nu }\omega
^{\mu } \\
\vec{R}^{\mu \nu } &=&\partial ^{\mu }\vec{\rho}^{\,\nu }-\partial ^{\nu }%
\vec{\rho}^{\,\mu } \\
F^{\mu \nu } &=&\partial ^{\mu }A^{\nu }-\partial ^{\nu }A^{\mu } \;.
\end{eqnarray}
The meson-nucleon couplings
$g_\sigma$, $g_\omega$, and $g_\rho$ are assumed to be vertex functions
of Lorentz-scalar bilinear forms of the nucleon field operators. In most
applications of the density-dependent hadron field theory these
couplings are chosen as functions of the vector density
$\rho _v = \sqrt{j_{\mu} j^{\mu}}$, with
$j_{\mu} = \bar{\psi} \gamma _{\mu} \psi$. Alternatively, the
couplings could be functionals of the scalar density
$\rho_s = \bar{\psi} \psi$. It has been shown, however, that
the vector density dependence produces better results for
finite nuclei~\cite{FLW.95}, and provides a more natural
relation between the self-energies of the density-dependent
hadron field theory and the Dirac-Brueckner microscopic
self-energies~\cite{HKL.01}. In the present work we choose
the vector density dependence for the meson-nucleon couplings.

The single-nucleon Dirac equation is derived by the variation of the
Lagrangian (\ref{lagrangian}) with respect to $\bar{\psi}$
\begin{equation}
i\partial _{t}\psi _{i} =
\left\{\bm{\alpha}\lbrack-i\bm{\nabla}-
{\bm V(r},t{\bm )]+}V({\bm r},t)+
{\bm \beta }\big(m+S({\bm r},t)\big) + \Sigma_0^R({\bm r},t)
\right\} \psi _{i} \;.
\label{dieq}
\end{equation}
The Dirac hamiltonian contains the scalar and vector 
nucleon self-energies defined by the following relations:
\begin{equation}
S({\bm r},t) = g_{\sigma }(\rho _v) \sigma({\bm r},t)
\end{equation}
\begin{equation}
V_{\mu}({\bm r},t)  = g_{\omega }(\rho _v) \omega _{\mu}({\bm r},t) +
g_{\rho } (\rho _v)\vec{\tau}\cdot \vec{\rho} _{\mu}({\bm r},t) +
e \frac{(1-\tau _{3})}{2} A _{\mu}({\bm r},t) \;.
\end{equation}
The density dependence of the vertex functions
$g_\sigma$, $g_\omega$, and $g_\rho$ produces the
{\it rearrangement} contribution to
the vector self-energy
\begin{equation}
\Sigma_0^R ({\bm r},t) =
  \frac {\partial g_{\omega }}{\partial \rho _v} \bar{\psi} \gamma ^{\nu }
  \psi \omega _{\nu} +
  \frac {\partial g_{\rho }}{\partial \rho _v} \bar{\psi} \gamma ^{\nu }
  \vec{\tau} \psi \cdot \vec{\rho} _{\nu} +
  \frac {\partial g_{\sigma }}{\partial \rho _v} \bar{\psi} \psi \sigma \; .
\end{equation}
The inclusion of the rearrangement self-energies
is essential for the energy-momentum conservation and the
thermodynamical consistency of the model~\cite{FLW.95,TW.99}.

In the time-dependent RMF model~\cite{VBR.95} one usually neglects the
retardation effects for the meson fields,  
and the self-energies are determined at each time by the
solutions of the Klein-Gordon and Poisson equations:
\begin{eqnarray}
\left[ -\Delta +m_{\sigma }^{2}\right] \,\sigma ({\bm r,}t) &=&-g_{\sigma
}(\rho _v)\,\rho _{s}({\bm r,}t) \nonumber \\
\left[ -\Delta +m_{\omega }^{2}\right] \,\omega _{\mu }({\bm r,}t)
&=&g_{\omega }(\rho _v)\,j_{\mu }({\bm r,}t)~,  \nonumber \\
\left[ -\Delta +m_{\rho }^{2}\right] \,\vec{\rho}_{\mu }({\bm r,}t)
&=&g_{\rho }(\rho _v)\,\,\vec{j}_{\mu }({\bm r},t)~,  \nonumber \\
-\Delta \,A_{\mu }({\bm r,}t) &=&e\,j_{c\mu }({\bm r},t)~.  
\label{KGeq}
\end{eqnarray}
This approximation is justified by the large masses
in the meson propagators. Retardation effects can be neglected
because of the short range of the corresponding meson exchange forces.
The explicit solutions of Eqs.~(\ref{KGeq}) read, respectively,
\begin{eqnarray}
\sigma ({\bm r},t) &=&-\int g_{\sigma } (\rho _v) D_{\sigma }({\bm r,
\bm r}^{\prime })\rho
_{s}({\bm r}^{\prime },t)d^{3}r^{\prime }~,  \nonumber \\
\omega _{\mu }({\bm r},t) &=&\int g_{\omega }(\rho _v) D_{\omega }({\bm r,
\bm r}^{\prime
})j_{\mu }({\bm r}^{\prime },t)d^{3}r^{\prime }~,  \nonumber \\
\vec{\rho}_{\mu }({\bm r},t) &=&\int g_{\rho }(\rho _v) D_{\rho }({\bm r,
\bm r}^{\prime })%
\vec{j}_{\mu }({\bm r}^{\prime },t)d^{3}r^{\prime }~,  \nonumber \\
A_{\mu }({\bm r},t) &=&e\int D_{c}({\bm r,\bm r}^{\prime })j_{c\mu }({\bm r}%
^{\prime },t)d^{3}r^{\prime }~\;,  
\label{KGsol}
\end{eqnarray}
with the Yukawa propagators 
\begin{equation}
D_{\phi }({\bm r},{\bm r}^{\prime })\,= \frac{1}{4\pi }\frac{\text{e}%
^{-m_{\phi }|{\bm r}-{\bm r}^{\prime }|}}{|{\bm r}-{\bm r}^{\prime }|}~,
\label{KGprop}
\end{equation}
where $\phi $ denotes the $\sigma $, $\omega $, and $\rho $ mesons, 
and the photon.

The sources of the fields 
are the nucleon densities and currents
calculated in the {\it no-sea} approximation
\begin{eqnarray}
\rho _{s}({\bm r},t) &=&\sum\limits_{i=1}^{A}\bar{\psi}_{i}^{{}}({\bm r}%
,t)\psi _{i}^{{}}({\bm r},t)~,  \nonumber \\
j_{\mu }({\bm r},t) &=&\sum\limits_{i=1}^{A}\bar{\psi}_{i}^{{}}({\bm r}%
,t)\gamma _{\mu }\psi _{i}^{{}}({\bm r},t)~,  \nonumber \\
\vec{j}_{\mu }({\bm r},t) &=&\sum\limits_{i=1}^{A}\bar{\psi}_{i}^{{}}({\bm r}%
,t)\vec{\tau}\gamma _{\mu }\psi _{i}^{{}}({\bm r},t)~,  \nonumber \\
j_{c\mu }({\bm r},t) &=&\sum\limits_{i=1}^{Z}\bar{\psi}_{i}^{{}}({\bm r}%
,t)\gamma _{\mu }\psi _{i}^{{}}({\bm r},t)~.  
\label{source}
\end{eqnarray}
where the summation is over all A occupied states in the Fermi sea,
i.e. only occupied single-nucleon states with positive energy
explicitly contribute to the nucleon self-energies. Even though
the stationary solutions for the negative-energy states
do not contribute to the densities in the {\it no-sea} approximation,
their contribution is implicitly included in 
the time-evolution of the nuclear system~\cite{Rin.01,VBR.95}.

The Relativistic Random Phase Approximation (RRPA)
represents the small amplitude limit of the
time-dependent relativistic mean-field theory. In the remainder of this
section we will derive the RRPA equations with density-dependent
meson-nucleon couplings from the response of the density matrix $\hat{\rho}(t)$
to an external field
\begin{equation}
\hat{F}(t)=\hat{F}\text{e}^{-i\omega t}+h.c.\;,  \label{extfi}
\label{ext}
\end{equation}
which oscillates with a small amplitude. In the single-particle
space this field is represented by the operator
\begin{equation}
\hat{f}(t) = \sum_{kl} \, f_{kl}(t) \; \hat{a}^{\dagger}_k \hat{a}^{}_l .
\end{equation}
The expression for the single-particle density matrix reads
\begin{equation}
\hat{\rho}({\bf r},{\bf r}^{\prime },t)=\sum\limits_{i=1}^{A}|\psi _{i}^{{}}(%
{\bf r},t)\rangle \langle \psi _{i}^{{}}({\bf r}^{\prime },t)|~.
\label{denmat}
\end{equation}
By writing the Dirac spinor in terms of large and small components
\begin{equation}
|\psi _{i}^{{}}({\bf r},t)\rangle =\left(
\begin{array}{c}
\,\,\,\,f_{i}({\bf r},t) \\
ig_{i}({\bf r},t)
\end{array}
\right) \;,   \label{dispi}
\end{equation}
the density matrix takes the form
\begin{equation}
\rho ({\bf r},{\bf r}^{\prime },t)=\left(
\begin{array}{cc}
\,\,\,\sum\limits_{i=1}^{A}f_{i}^{{}}({\bf r},t)f_{i}^{\dagger }({\bf r}%
^{\prime },t) & -i\sum\limits_{i=1}^{A}f_{i}^{{}}({\bf r},t)g_{i}^{\dagger }(%
{\bf r}^{\prime },t) \\
i\sum\limits_{i=1}^{A}g_{i}^{{}}({\bf r},t)f_{i}^{\dagger }({\bf r}^{\prime
},t) & \,\,\,\,\,\,\sum\limits_{i=1}^{A}g_{i}^{{}}({\bf r},t)g_{i}^{\dagger
}({\bf r}^{\prime },t)
\end{array}
\right) \quad \;.  \label{denmat1}
\end{equation}
The equation of motion for the density operator reads
\begin{equation}
i\partial _{t}\hat{\rho}=\left[ \hat{h}(\hat{\rho})+\hat{f}(t),\hat{\rho}%
\right] \;,  \label{eqmot}
\end{equation}
and in the small amplitude limit the density matrix is expanded
to linear order
\begin{equation}
\hat{\rho}(t)=\hat{\rho}^{(0)}+\delta \hat{\rho}(t) \;,  \label{linres}
\end{equation}
where $\hat{\rho}^{(0)}$ is the stationary ground-state density.
From the definition of the density matrix (\ref{denmat}), it follows
that $\hat{\rho}\left( t\right) $ is a projector at all times, i.e.
$\hat{\rho}\left( t\right) ^{2}=$ $\hat{\rho}\left( t\right) $.
In particular, this means that the eigenvalues of $\hat{\rho}^{(0)}$
are 0 and 1. In the non-relativistic case particle states
above the Fermi level correspond to the eigenvalue 0, and hole states
in the Fermi sea correspond to the eigenvalue 1. In the relativistic
case, one also has to take into account states from the Dirac sea.
In the {\it no-sea} approximation these states are not occupied,
i.e. they correspond to the eigenvalue 0 of the density matrix.
\begin{equation}
\rho _{kl}^{(0)}=\delta _{kl}\rho _{k}^{(0)}=\left\{
\begin{array}{ll}
0 & \text{for unoccupied states above the Fermi level (index }p\text{)}
\\
1 & \text{for occupied states in the Fermi sea (index }h\text{)\quad } \\
0 & \text{for unoccupied states in the Dirac sea (index }\alpha \text{)}\; .
\end{array}
\right.   \label{occup}
\end{equation}
$\hat{\rho}\left( t\right) ^{2}=$ $\hat{\rho}\left( t\right) $ also
implies, in leading order,
\begin{equation}
\hat{\rho}^{(0)}\delta \hat{\rho}+\delta \hat{\rho}\hat{\rho}^{(0)}=\delta
\hat{\rho}~\;,  
\end{equation}
and this means that the only
non-vanishing matrix elements of $\delta \hat{\rho}$ are:
$\delta \rho _{ph}$, $\delta \rho_{hp}$,
$\delta \rho _{\alpha h}$, and $\delta \rho _{h\alpha }$. 
These are
determined by the solution of equation (\ref{eqmot}), which in
the linear approximation reads
\begin{equation}
i\partial _{t}\delta \hat{\rho}=\left[ \hat{h}^{(0)},\delta \hat{\rho}\right]
+\left[ \frac{\partial \hat{h}}{\partial \rho }\delta \rho ,\hat{\rho}^{(0)}%
\right] +\left[ \hat{f},\hat{\rho}^{(0)}\right] \;,  
\end{equation}
with
\begin{equation}
\frac{\partial \hat{h}}{\partial \rho }\delta \rho =\sum_{ph}\frac{\partial
\hat{h}}{\partial \rho _{ph}}\delta \rho _{ph}+\frac{\partial \hat{h}}{%
\partial \rho _{hp}}\delta \rho _{hp}+\sum_{\alpha h}\frac{\partial \hat{h}}{%
\partial \rho _{\alpha h}}\delta \rho _{\alpha h}+\frac{\partial \hat{h}}{%
\partial \rho _{h\alpha }}\delta \rho _{h\alpha } \;.  
\end{equation}
Under the influence of the external field (\ref{ext}), 
in the small amplitude limit $\delta \rho$ also exhibits
the harmonic time dependence e$^{-i\omega t}$. Taking into account 
that $\hat{h}_{kl}^{(0)}=\delta _{kl}\epsilon
_{k}$ is diagonal in the stationary basis, the resulting 
RRPA equations read 
\begin{eqnarray}
(\omega -\epsilon _{p}+\epsilon _{h})\delta \rho _{ph}
&=&f_{ph}+\sum_{p^{\prime }h^{\prime }}V_{ph^{\prime }hp^{\prime }}\delta
\rho _{p^{\prime }h^{\prime }}+V_{pp^{\prime }hh^{\prime }}\delta \rho
_{h^{\prime }p^{\prime }}+\sum_{\alpha^{\prime }h^{\prime }}V_{ph^{\prime
}h\alpha^{\prime }}\delta \rho _{\alpha^{\prime }h^{\prime }}+
V_{p\alpha^{\prime }hh^{\prime }}
\delta \rho_{h^{\prime }\alpha^{\prime }}
\nonumber
\\
(\omega -\epsilon_{\alpha }+\epsilon _{h})\delta \rho _{\alpha h}
&=&f_{\alpha h}+\sum_{p^{\prime }h^{\prime }}V_{\alpha h^{\prime }hp^{\prime
}}\delta \rho _{p^{\prime }h^{\prime }}+V_{\alpha p^{\prime }hh^{\prime
}}\delta \rho _{h^{\prime }p^{\prime }}+\sum_{\alpha ^{\prime }h^{\prime
}}V_{\alpha h^{\prime }h\alpha ^{\prime }}\delta \rho _{\alpha ^{\prime
}h^{\prime }}+V_{\alpha \alpha ^{\prime }hh^{\prime }}\delta \rho
_{h^{\prime }\alpha ^{\prime }}  \nonumber \\
(\omega -\epsilon _{h}+\epsilon _{p})\delta \rho _{hp}
&=&f_{hp}+\sum_{p^{\prime }h^{\prime }}V_{hh^{\prime }pp^{\prime }}\delta
\rho _{p^{\prime }h^{\prime }}+V_{hp^{\prime }ph^{\prime }}\delta \rho
_{h^{\prime }p^{\prime }}+\sum_{\alpha ^{\prime }h^{\prime }}V_{hh^{\prime
}p\alpha ^{\prime }}\delta \rho _{\alpha ^{\prime }h^{\prime }}+V_{h\alpha
^{\prime }ph^{\prime }}\delta \rho _{h^{\prime }\alpha ^{\prime }}
\nonumber
\\
(\omega -\epsilon _{h}+\epsilon_{\alpha })\delta \rho _{h\alpha }
&=&f_{h\alpha }+\sum_{p^{\prime }h^{\prime }}V_{hh^{\prime }
\alpha p^{\prime}}\delta \rho _{p^{\prime }h^{\prime }}+
V_{hp^{\prime }\alpha h^{\prime}}
\delta \rho _{h^{\prime }p^{\prime }}+
\sum_{\alpha ^{\prime }h^{\prime}}
V_{hh^{\prime }\alpha \alpha ^{\prime }}
\delta \rho _{\alpha ^{\prime}h^{\prime }}+
V_{h\alpha ^{\prime }\alpha h^{\prime }}
\delta \rho_{h^{\prime }\alpha ^{\prime }}
\nonumber 
\\
\label{rpaeq}
\end{eqnarray}
or, in matrix form
\begin{equation}
\left[ \omega \left(
\begin{array}{cc}
1 & 0 \\
0 & -1
\end{array}
\right) -\left(
\begin{array}{cc}
A & B \\
B^{\ast } & A^{\ast }
\end{array}
\right) \right] \left(
\begin{array}{c}
X \\
Y
\end{array}
\right) =\left(
\begin{array}{c}
F \\
\bar{F}
\end{array}
\right) \;.  
\end{equation}
\bigskip 

The RRPA matrices $A$ and $B$ read
\begin{eqnarray}
A &=&\left(
\begin{array}{cc}
(\epsilon _{p}-\epsilon _{h})\delta _{pp^{\prime }}\delta _{hh^{\prime }} &
\\
& (\epsilon _{\alpha }-\epsilon _{h})\delta _{\alpha \alpha ^{\prime
}}\delta _{hh^{\prime }}
\end{array}
\right) +\left(
\begin{array}{cc}
V_{ph^{\prime }hp^{\prime }} & V_{ph^{\prime }h\alpha ^{\prime }} \\
V_{\alpha h^{\prime }hp^{\prime }} & V_{\alpha h^{\prime }h\alpha ^{\prime }}
\end{array}
\right)   \label{E3.10} \\
B &=&\left(
\begin{array}{cc}
V_{pp^{\prime }hh^{\prime }} & V_{p\alpha ^{\prime }hh^{\prime }} \\
V_{\alpha p^{\prime }hh^{\prime }} & V_{\alpha \alpha ^{\prime }hh^{\prime }}
\end{array}
\right)   \label{rpamat}
\end{eqnarray}
and the amplitudes $X$ and $Y$ are defined
\begin{equation}
X=\left(
\begin{array}{c}
\delta \rho _{ph} \\
\delta \rho _{\alpha h}
\end{array}
\right) ,\quad Y=\left(
\begin{array}{c}
\delta \rho _{hp} \\
\delta \rho _{h\alpha }
\end{array}
\right) ~.  \label{rpaamp}
\end{equation}
The vectors which represent the external field contain
the matrix elements
\begin{equation}
F=\left(
\begin{array}{c}
f_{ph} \\
f_{\alpha h}
\end{array}
\right) ,\quad \bar{F}=\left(
\begin{array}{c}
f_{hp} \\
f_{h\alpha }
\end{array}
\right) ~.  \label{extvec}
\end{equation}
In the self-consistent RRPA the matrix elements of the residual interaction 
are derived 
from the Dirac hamiltonian of Eq. (\ref{dieq}):
\begin{eqnarray}
V_{abcd} &=& \frac{\partial h_{ac}}{\partial \rho_{db}} \nonumber \\
        &=& \int \Psi_a^+(\bm r_1) \Psi_b^+(\bm r_2) V(\bm r_1,\bm r_2)
	      \Psi_c(\bm r_1) \Psi_d(\bm r_2) d^3 r_1 d^3 r_2 \;.
\end{eqnarray}	
In order to calculate the contributions of each meson field to 
$V(\bm r_1,\bm r_2)$, we expand the meson-nucleon couplings 
and their derivatives around
the ground-state density $\rho_v^0$
\begin{eqnarray}
g_i(\rho_v) &=& g_i(\rho_v^{0})+
\frac{\partial g_i}{\partial\rho_v}\Bigg{|}_{0}\delta\rho_v  \nonumber \\
\frac{\partial g_i}{\partial \rho_v}&=&\frac{\partial g_i}{\partial \rho_v}
\Bigg{|}_{0}+ \frac{\partial^2 g_i}{\partial \rho_v^2}\Bigg{|}_{0}
\delta\rho_v \quad \;.
\label{lin}
\end{eqnarray}	        		  
If for the meson fields appearing in the scalar and vector nucleon
self-energies we use the explicit solutions (\ref{KGsol}) in terms
of the meson propagators and nucleon densities and currents, the 
individual contributions of the meson fields to $V(\bm r_1,\bm r_2)$ 
are obtained from the
particle-hole matrix element of the Dirac hamiltonian:
\begin{itemize}
\item the contribution of the isoscalar-scalar sigma field 
\begin{eqnarray}
V_{\sigma}(\bm r_1, \bm r_2)&=& - \beta_1\beta_2 
g_{\sigma}(\rho_v(\bm r_1))g_{\sigma}(\rho_v(\bm r_2)) 
D_{\sigma}(\bm r_1,\bm r_2) \nonumber \\
&-&\Bigg\{ \beta_1 \openone_2 
     \frac{\partial g_{\sigma}}{\partial \rho_v(\bm r_1)} + 
     \openone_1\openone_2
     \frac{\partial^2 g_{\sigma}}{\partial \rho_v^2(\bm r_1)}\rho_s(\bm r_1)
     +\openone_1\beta_2\frac{\partial g_{\sigma}}{\partial 
     \rho_v(\bm r_1)} \Bigg\}
     \frac{I_{\sigma}(r_1)}{r_1}\delta(\bm r_1-\bm r_2) \nonumber \\
&-&\Bigg\{ \beta_1 \openone_2 g_{\sigma}
     (\rho_v(\bm r_1))\frac{\partial g_{\sigma}}{\partial 
     \rho_v(\bm r_2)}\rho_s(\bm r_2) 
     + \openone_1\beta_2 \frac{\partial g_{\sigma}}{\partial \rho_v(\bm r_1)}
        \rho_s(\bm r_1) g_{\sigma}(\rho_v(\bm r_2))\nonumber \\ 
&+& \openone_1\openone_2
      \frac{\partial g_{\sigma}}{\partial \rho_v(\bm r_1)}\rho_s(\bm r_1)
       \frac{\partial g_{\sigma}}
       {\partial \rho_v(\bm r_2)}\rho_s(\bm r_2) \Bigg\}
       D_{\sigma}(\bm r_1,\bm r_2) \;, 
\label{ress}
\end{eqnarray} 
where
\begin{displaymath}
I_{\sigma}(r_1) = \int rg_{\sigma}(\rho_v(r))
D_{\sigma}^{0}(r_1,r)\rho_s(r)dr \; ;  
\end{displaymath}
\item the contribution of the isoscalar-vector omega field 
\begin{eqnarray}
V_{\omega}(\bm r_1, \bm r_2)&=& (\beta \gamma^{\mu})_1
(\beta \gamma_{\mu})_2 g_{\omega}(\rho_v(\bm r_1))
   g_{\omega}(\rho_v(\bm r_2)) D_{\omega}(\bm r_1,\bm r_2)\nonumber \\
&+& \Bigg\{ 
    2 \frac{\partial g_{\omega}}{\partial \rho_v(\bm r_1)} + 
     \frac{\partial^2 g_{\omega}}
     {\partial \rho_v^2(\bm r_1)}\rho_v(\bm r_1) \Bigg\}
	\openone_1\openone_2
     \frac{I_{\omega}(r_1)}{r_1}\delta(\bm r_1-\bm r_2) \nonumber \\
&+&\Bigg\{ g_{\omega}(\rho_v(\bm r_1))
     \frac{\partial g_{\omega}}{\partial \rho_v(\bm r_2)}\rho_v(\bm r_2)+
    \frac{\partial g_{\omega}}{\partial \rho_v(\bm r_1)}\rho_v(\bm r_1)
    g_{\omega}(\rho_v(\bm r_2))\nonumber\\
    &+&  \frac{\partial g_{\omega}}{\partial \rho_v(\bm r_1)}\rho_v(\bm r_1)
       \frac{\partial g_{\omega}}
       {\partial \rho_v(\bm r_2)}\rho_v(\bm r_2) \Bigg\}
       \openone_1 \openone_2 D_{\omega}(\bm r_1,\bm r_2) \;,
\label{reso}
\end{eqnarray}  
where
\begin{displaymath}
I_{\omega}(r_1) = \int rg_{\omega}(\rho_v(r))
D_{\omega}^{0}(r_1,r)\rho_v(r)dr \; ;
\end{displaymath}
\item the contribution of the isovector-vector rho field
\begin{eqnarray}
V_{\rho}^1(\bm r_1, \bm r_2)&=&(\beta \gamma^{\mu})_1(\beta \gamma_{\mu})_2 
\tau_1^3 \tau_2^3 
   g_{\rho}(\rho_v(\bm r_1))g_{\rho}(\rho_v(\bm r_2)) 
   D_{\rho}(\bm r_1,\bm r_2)\nonumber \\
&+&\Bigg\{ 
     \frac{\partial g_{\rho}}{\partial \rho_v(\bm r_1)}\tau_1^3 + 
     \frac{\partial g_{\rho}}{\partial \rho_v(\bm r_1)}\tau_2^3 +
     \frac{\partial^2 g_{\rho}}{\partial \rho_v^2(\bm r_1)}
     \rho_{tv}(\bm r_1) \Bigg\}
	\openone_1\openone_2
     \frac{I_{\rho}(r_1)}{r_1}\delta(\bm r_1-\bm r_2) \nonumber \\
&+&\Bigg\{ g_{\rho}(\rho_v(\bm r_1))\tau_1^3
     \frac{\partial g_{\rho}}{\partial \rho_v(\bm r_2)}\rho_{tv}(\bm r_2)+
    \frac{\partial g_{\rho}}{\partial \rho_v(\bm r_1)}\rho_{tv}(\bm r_1)
    g_{\rho}(\rho_v(\bm r_2))\tau_2^3\nonumber\\
     &+& \frac{\partial g_{\rho}}{\partial \rho_v(\bm r_1)}\rho_{tv}(\bm r_1)
       \frac{\partial g_{\rho}}
       {\partial \rho_v(\bm r_2)}\rho_{tv}(\bm r_2) \Bigg\}
       \openone_1 \openone_2 D_{\rho}(\bm r_1,\bm r_2) \; ,
\label{resr}
\end{eqnarray}  
where
\begin{displaymath}
I_{\rho}(r_1) = \int rg_{\rho}(\rho_v(r))D_{\rho}^{0}(r_1,r)\rho_{tv}(r)dr \; ;
\end{displaymath}         
\item and finally the contribution of the Coulomb field
\begin{eqnarray}
V_c(\bm{r}_1, \bm{r}_2)=e^2(\beta \gamma^{\mu})_1(\beta \gamma_{\mu})_2
               D_c(\bm{r}_1, \bm{r}_2) \; .
\label{resc}
\end{eqnarray}	
\end{itemize}
the subscripts 1 and 2 of the Dirac matrices refer to particle 1 and 2,
respectively. $\rho_v$, $\rho_s$ and $\rho_{tv}$ denote the vector, scalar,
and isovector-vector density, respectively, and the derivatives of the 
meson-nucleon couplings with respect to the vector density are evaluated 
at ground-state density $\rho_v^0$.
The radial integrals $I_{\phi}(r)$ ($\phi\equiv \sigma, \omega, \rho$)
contain $D_{\phi}^{0}(r,r^{\prime})$, which is the radial factor
in the first term of the multipole expansion of the Yukawa 
propagator (\ref{KGprop}):
\begin{equation}
D_{\phi}({\bm r},{\bm r}^{\prime}) = \frac{1}{r r^{\prime}}
\sum_{L=0}^\infty D_{\phi}^{L}(r,r^{\prime})
\sum_{M=-L}^L Y_{LM}(\Omega) Y^{*}_{LM}(\Omega^{\prime})
\end{equation}

We notice that, in addition to the direct contribution of the 
meson exchange interactions (first terms in Eqs. (\ref{ress})-(\ref{resr})),
the explicit density dependence of the meson-nucleon couplings 
introduces a number of {\it rearrangement} terms in the residual 
two-body interaction $V({\bm r}_1,{\bm r}_2)$. These
rearrangement terms are essential for fully consistent RRPA calculations.
Only when their contribution is included in the matrix elements
of the residual interaction, it becomes possible to reproduce reasonably
well the excitation energies of giant multipole resonances. Without
rearrangement terms, one finds discrepancies of the order of several MeV
between the experimental excitation energies and the RRPA peak energies,
calculated with relativistic effective interactions that are adjusted
to ground state properties of spherical nuclei. A very similar effect
is observed in RRPA calculations based on effective forces 
with non-linear meson self-interactions, when the 
contribution of the non-linear terms is
not included in the matrix elements of the residual 
interaction~\cite{Mawa.01}.


\section{\label{secIII}Illustrative RRPA calculations: Giant Resonances}

In this section the RRPA with density-dependent meson-nucleon
couplings is applied 
in illustrative calculations of giant resonances in spherical
nuclei. In particular, we analyze the isoscalar monopole, the 
isovector dipole and the isoscalar quadrupole resonances in $^{208}$Pb.
We will show which isoscalar and isovector properties of the effective
mean-field interactions affect the 
multipole strength distributions, and how the results
of RRPA calculations can be used to constrain the effective 
interaction. 

For the density dependence of the meson-nucleon couplings we adopt the
functionals used in Refs.\cite{TW.99,HKL.01,NVFR.02}. 
The coupling of the $\sigma$-meson and $\omega$-meson to the nucleon
field reads
\begin{equation}
g_i(\rho) = g_i(\rho_{\rm sat}) f_i(x)\quad {\rm for}\quad i=\sigma, \omega\;,
\label{coupl}
\end{equation}
where
\begin{equation}
f_i(x) = a_i \frac{1 + b_i(x+d_i)^2}{1 + c_i(x+d_i)^2}
\label{func}
\end{equation}
is a function of $x = \rho /\rho_{\rm sat}$, and $\rho_{\rm sat}$ denotes
the baryon density at saturation in symmetric nuclear matter. 
The eight real parameters
in (\ref{func}) are not independent. The five constraints $f_i(1)=1$,
$f_\sigma^{\prime\prime}(1) = f_\omega^{\prime\prime}(1)$, and
$f_i^{\prime\prime}(0)=0$, reduce the number of independent parameters
to three. Three additional parameters in the isoscalar channel are:
$g_\sigma(\rho_{\rm sat})$, $g_\omega(\rho_{\rm sat})$, and
$m_\sigma$ - the mass of the phenomenological sigma-meson. 
For the $\rho$-meson coupling the functional
form of the density dependence is
suggested by DB calculations of asymmetric nuclear matter~\cite{JL.98}
\begin{equation}
\label{drho}
g_{\rho}(\rho) = g_{\rho}(\rho_{\rm sat})~{\rm exp}
\left [-a_{\rho} (x-1)\right ]\;.
\label{grho}
\end{equation}
The isovector channel is parameterized by $g_{\rho}(\rho_{\rm sat})$ and
$a_{\rho}$. For the masses of the $\omega$ and $\rho$ mesons 
usually the free values are used: $m_\omega = 783$ MeV and 
$m_\rho = 763$ MeV. In principle one could also consider the 
density dependence of the meson masses. However, since the effective
meson-nucleon coupling in nuclear matter 
is determined by the ratio $g/m$, the choice of a
phenomenological density dependence of the couplings makes an
explicit density dependence of the masses redundant. 

The eight independent parameters, seven coupling parameters and
the mass of the $\sigma$-meson, are adjusted to reproduce the
properties of symmetric and asymmetric nuclear matter,
binding energies, charge radii and neutron radii of spherical nuclei.
In particular, in Ref.~\cite{NVFR.02} we have introduced the 
density-dependent meson-exchange effective interaction
(DD-ME1), whose parameters are displayed in Table~\ref{tab1}.
The seven coupling parameters and the $\sigma$-meson mass have been
simultaneously adjusted to properties of symmetric and asymmetric
nuclear matter, and to ground-state properties (binding energies,
charge radii and differences between neutron and proton radii)
of twelve spherical nuclei. For the
open shell nuclei pairing correlations have
been treated in the BCS approximation with empirical pairing gaps
(five-point formula). 

In Ref.~\cite{NVFR.02} the relativistic Hartree-Bogoliubov (RHB) model
with the density-dependent interaction DD-ME1 in the
$ph$-channel, and with the finite range Gogny interaction D1S in
the $pp$-channel, has been tested
in the analysis of the equations of state for symmetric
and asymmetric nuclear matter, and of ground-state properties of the
Sn and Pb isotopic chains. It has been shown that, as compared to
standard non-linear relativistic mean-field effective forces,
the interaction DD-ME1 has better isovector properties and therefore
provides an improved description of asymmetric nuclear matter, neutron
matter and nuclei far from stability.

In the present analysis we perform fully consistent RRPA calculations
of isoscalar monopole, isovector dipole and isoscalar quadrupole 
giant resonances in $^{208}$Pb. The single-particle basis and the 
particle-hole couplings are obtained from the same effective Lagrangian,
and the configuration space includes both particle-hole pairs, as
well as pairs formed from hole states and negative-energy states from
the Dirac sea. Our starting point is the DD-ME1
effective force, both in the Dirac hamiltonian (\ref{dieq}), as well
as the residual interaction. We then proceed to construct families of
density-dependent interactions with some given characteristic 
(compressibility, asymmetry energy, etc.), and study the resulting 
properties of giant resonances. 

The isoscalar giant monopole resonance (ISGMR) represents the most simple 
mode of collective excitations in nuclei. In particular, the ISGMR in 
heavy nuclei is the only source of experimental information on 
the nuclear matter compression modulus $K_\infty$. This quantity determines
basic properties of nuclei, supernovae explosions, neutron stars
and heavy-ion collisions. The range of values of $K_\infty$ has 
been deduced from the measured excitation energies of the ISGMR
in spherical nuclei. The presently available
experimental data set, however, does not limit
the range of $K_{\infty}$ to better than $200 - 300$ MeV.
The microscopic determination of the nuclear matter compressibility 
is based on the construction of sets of effective interactions
that differ mostly by their prediction of the excitation energies
of ISGMR, i.e. by the value of $K_{\infty}$, but otherwise
reproduce reasonably well experimental data on ground-state nuclear
properties~\cite{Bla.80,Bla.95}. Effective interactions with different
values of $K_{\infty}$ are used to calculate bulk ground-state
properties of heavy spherical nuclei in a self-consistent mean-field
framework, and RPA or time-dependent mean-field calculations are
performed for the isoscalar monopole excitations. Such a fully
consistent calculation of both ground-state properties, as well
as ISGMR excitation energies, restricts the range of possible
values for $K_{\infty}$. However, since there are also other 
effects beyond the mean-field level which influence the isoscalar
monopole resonance
(anharmonicities, pairing, coupling between single-nucleon
and collective motion), it has been argued ~\cite{Bla.95}
that, rather than on
the systematics over the whole periodic table, the
determination of the nuclear compressibility should rely more on
a good measurement and microscopic calculations of GMR in a
single heavy nucleus such as $^{208}$Pb. Microscopic calculations
have been performed both in the non-relativistic and in the 
relativistic mean-field framework. Modern 
non-relativistic Hartree-Fock plus RPA calculations, using both
Skyrme and Gogny effective interactions, indicate that the value of
$K_{\infty}$ should be in the range 210-220 MeV~\cite{Bla.80,Bla.95}. 
In particular, in Ref.~\cite{Bla.95} a set of effective Gogny forces
was generated, which on one hand allow a good description of
static properties of nuclei, and on the other hand span the 
range $210 \leq K_{\infty} \leq 300$ MeV. It was shown that
the RPA calculations reproduce the available experimental data on
ISGMR in medium-heavy and heavy nuclei only for 
$K_{\infty}$ in the range 210-220 MeV.
In Ref.~\cite{Far.97} it has been shown that even generalized Skyrme
forces, with both density- and momentum-dependent terms, can only reproduce
the measured breathing mode energies for values of $K_{\infty}$ in the 
range $215\pm 15$ MeV.
In relativistic mean-field models based on non-linear meson 
self-interactions on the other hand, results of both RRPA and
time-dependent calculations suggest that
empirical GMR energies are best reproduced by an effective
force with $K_{\infty}\approx 250 - 270$ MeV~\cite{Vre.97,MGW.01}.
It has to be emphasized, however, that even though relativistic 
calculations have been performed using non-linear effective interactions 
with different values of $K_{\infty}$, these forces were not constructed
specifically with the purpose of determining $K_{\infty}$. Rather,
standard non-linear effective interactions have been used, which
also exhibit other 
differences that could affect the microscopic determination of the nuclear
matter compressibility.

Starting from DD-ME1, in this work we have generated a consistent set of
relativistic density-dependent effective interactions with 
$220 \leq K_{\infty} \leq 280$ MeV. The same functional form for the
density dependence for the meson-nucleon couplings has been 
used for these forces and, except for the value of $K_{\infty}$,
their parameters have been adjusted to the same set of experimental
data on ground-state properties of twelve spherical nuclei~\cite{NVFR.02}.
The results of fully consistent RRPA calculations with these
forces are shown in Fig.~\ref{figA}, where we display the calculated
excitation energies of ISGMR in $^{208}$Pb as function of the 
nuclear matter compressibility. The shaded region denotes the 
range of presently available experimental data~\cite{YCL.99}.
We notice that, in accordance with the results obtained with 
relativistic effective forces with non-linear meson self-interactions,
only the density-dependent interactions with 
$K_{\infty}\approx 260 - 270$ MeV reproduce the experimental value. 
We have also verified that ISGMR excitation energies for lighter nuclei, 
calculated with these particular interactions, are closest to the 
empirical curve $E_x \approx 80~A^{-1/3}$ MeV and that they reproduce
the experimental excitation energies. For the
density-dependent effective interaction with $K_\infty = 270$ MeV,
in Fig.~\ref{figB} we display the isoscalar monopole
strength distribution and transition densities in $^{208}$Pb.
The position of the ISGMR peak is at $E = 14.1$ MeV, and we plot 
the proton, neutron and total isoscalar transition 
densities.

The present analysis, therefore, confirms that there is a pronounced
difference between the values of the 
nuclear matter compression modulus predicted 
by microscopic non-relativistic ($K_{\infty} \approx 210-230$ MeV) 
and relativistic ($K_{\infty} \approx 250-270$ MeV) 
mean-field plus random phase approximation calculations. The origin
of this discrepancy is at present not understood, even though there
are some indications that it might be due, at least in part,
to the differences in the density dependence of the asymmetry energy
predicted by non-relativistic and relativistic models~\cite{Pie.02}.

The calculated properties of isovector dipole giant resonances (IVGDR)
will be predominantly determined by the isovector channel of the 
effective interaction. In particular, the excitation energies of IVGDR
can be directly related to the nuclear matter asymmetry energy.
The energy per particle of asymmetric nuclear
matter can be expanded about the equilibrium density $\rho_{\rm sat}$
in a Taylor series in $\rho$ and $\alpha$~\cite{Lee.98}
\begin{equation}
E(\rho,\alpha) = E(\rho,0) + S_2(\rho) \alpha^2 + S_4(\rho) \alpha^4 + \cdots
\label{taylor}
\end{equation}
where
\begin{equation}
\alpha \equiv \frac{N-Z}{N+Z} \;.
\end{equation}

\begin{equation}
E(\rho,0) = - a_v + \frac{K_0}{18 \rho_{\rm sat}^2}~
(\rho - \rho_{\rm sat})^2 + ...
\end{equation}
and
\begin{equation}
S_2(\rho) = a_4 + \frac{p_0}{\rho_{\rm sat}^2}~(\rho - \rho_{\rm sat}) +
\frac{\Delta K_0}{18 \rho_{\rm sat}^2}~ (\rho - \rho_{\rm sat})^2 + \cdots
\label{S2}
\end{equation}
The empirical value of the asymmetry energy 
at saturation density (volume asymmetry) $S_2(\rho_{\rm sat}) = a_4 =
30\pm 4$ MeV. The parameter $p_0$ defines the linear density dependence
of the asymmetry energy, and $\Delta K_0$ is the correction to the
incompressibility. The contribution of the term $S_4(\rho) \alpha^4$
in (\ref{taylor}) is very small
in ordinary nuclei and the coefficient is not constrained in the mean-field
approximation.

A ground-state nuclear property which is directly determined by the 
asymmetry energy is the difference between the neutron and the proton radii.
In a recent study of neutron
radii in non-relativistic and covariant mean-field models~\cite{Fur.01},
the linear correlation between the neutron skin
and the symmetry energy has been analyzed. 
In particular, the analysis has shown that there is a
very strong linear correlation between the neutron skin thickness in
$^{208}$Pb and the individual parameters that determine the
symmetry energy $S_2(\rho)$: $a_4$, $p_0$ and $\Delta K_0$. The empirical
value of $r_n - r_p$ in $^{208}$Pb ($0.20 \pm 0.04$ fm from proton
scattering data~\cite{SH.94}, and $0.19 \pm 0.09$ fm from the alpha scattering
excitation of the isovector giant dipole resonance~\cite{Kra.94}) places the
following constraints on the values of the parameters of the symmetry
energy: $a_4 \approx 30-34$ MeV, 2 Mev/fm$^3 \leq p_0 \leq$ 4 Mev/fm$^3$,
and $-200$ MeV $\leq \Delta K_0 \leq -50$ MeV. 

Properties of isovector collective modes in finite nuclei should, 
in principle, provide additional constraints on the isovector channel
of the effective interaction. In an analysis of Skyrme forces and
giant resonances in exotic nuclei~\cite{Rei.99}, Reinhard noticed 
a somewhat surprising property of the IVGDR: while it is true that
the excitation energy of this resonance is sensitive to the 
volume asymmetry $a_4$, the resonance energy decreases by
increasing the asymmetry energy at saturation. This was qualitatively 
explained by noticing that an increase in the volume asymmetry is 
always accompanied by an increase of the slope $p_0$, i.e. of the
linear density dependence of the asymmetry energy. In order to study 
this effect in a more quantitative way, we have generated, starting 
from DD-ME1, a set of eight density dependent effective interactions with
30 MeV $\leq a_4 \leq$ 37 MeV. The parameters of the density-dependent
meson-nucleon couplings have been adjusted in such a way that,
while increasing $a_4$ in units of 1 MeV, the resulting   
effective interactions still reproduce the same set of 
data on ground-state properties of spherical nuclei, that was used
for the original interaction DD-ME1~\cite{NVFR.02}. This means that
these effective interactions essentially differ only in their description
of the asymmetry energy curve as function of the baryon density.

The resulting nuclear matter asymmetry energy curves, and the calculated
IVGDR excitation energies in $^{208}$Pb, are displayed in Fig.~\ref{figC}.
In the upper left panel we plot the RRPA excitation energy of the
IVGDR in $^{208}$Pb as function of the volume asymmetry $a_4$.
Similar to what has been observed in Ref.~\cite{Rei.99}, the 
resonance energy decreases with increasing $a_4$. The reason for this
decrease is shown in the lower left panel, where we plot the 
corresponding values of the slope parameter $p_0$, which 
defines the linear density dependence of the asymmetry energy. 
We notice that, in order to reproduce the bulk properties of spherical
nuclei, an increase of $a_4$ necessitates a non-linear increase of $p_0$.
The resulting asymmetry energy curves as functions of the baryon density 
are shown in the right panel of Fig.~\ref{figC}. The increase of 
$p_0$ with $a_4$ implies a transition from a parabolic to an 
almost linear density dependence of $S_2$ in the density region 
$\rho \leq 0.2$ fm$^{-3}$. This means, in particular, that the increase
of the asymmetry energy at saturation point will produce an 
effective decrease of $S_2$ below $\rho \approx 0.1$ fm$^{-3}$. But this is,
of course, the density region characteristic for the IVGDR. We find,
therefore, that the excitation energy of the IVGDR decreases with 
increasing $S_2(\rho_{\rm sat}) \equiv a_4$, because this increase
implies a decrease of $S_2$ at low densities characteristic for the 
surface modes. In the upper left panel of Fig.~\ref{figC} we also
compare the calculated IVGDR peak energies for $^{208}$Pb, with the 
experimental value of $13.3\pm 0.1$~\cite{Rit.93}. It appears that
the experimental IVGDR excitation energy constrains 
the nuclear matter asymmetry energy at saturation density to the 
interval 34 MeV $\leq a_4 \leq$ 36 MeV. For the effective interaction 
with $a_4 = 35$ MeV, in Fig.~\ref{figD} we display the RRPA 
isovector dipole strength
distribution and the corresponding proton, neutron, and
total isovector transition densities for the peak at 13.3 MeV in $^{208}$Pb.

Fig.~\ref{figE} illustrates what happens when the increase of the 
nuclear matter asymmetry energy at saturation density is not accompanied
by an increase of the slope parameter $p_0$. Starting with DD-ME1, which 
has $a_4 = 33.1$ MeV, we have generated a set of effective interactions 
with different values of $a_4$, but now they all have the same slope parameter
$p_0$ (lower left panel), i.e. the parameters are not readjusted 
to reproduce the data set of ground state properties of spherical nuclei.   
Binding energies and radii are only approximately reproduced 
with these effective interactions.
The resulting asymmetry energy curves as functions of the baryon density
are plotted in the right panel. Since $p_0$ is constant, 
by increasing $a_4$ the asymmetry
energy $S_2$ increases for all densities. As a result,
the IVGDR peak energies of $^{208}$Pb increase linearly with $a_4$
(upper left panel). 

In non-relativistic RPA calculations, the excitation energy 
of the isoscalar giant quadrupole resonance (ISGQR) can be directly related
to the nucleon effective mass that characterizes a given effective 
interaction. In the non-relativistic mean-field approximation, 
the total effective
mass $m^*$ of a nucleon in a nucleus characterizes the energy
dependence of an effective local potential that is equivalent
to the, generally nonlocal and frequency dependent, microscopic
nuclear potential~\cite{MS.96}. $m^*$ is a measure of the density 
of single-nucleon states around the Fermi surface and, therefore, 
it affects the giant resonances. For Skyrme interactions, in particular, 
a linear dependence on $m^*$ is found for the RPA excitation energies 
of the ISGQR. The larger the effective mass, i.e. the 
higher the density of states around the Fermi surface, the lower
is the calculated ISGQR excitation energy. Both the calculation of
ground-state properties in spherical nuclei, as well as the RPA results
for ISGQR excitation energies, place the following constraint on the
nucleon effective mass for Skyrme-type interactions:
$m^*/m = 0.8 \pm 0.1$~\cite{Rei.99}.

In the relativistic framework the concept of effective nucleon mass
is more complicated. In addition to the $k-mass$ (characterizes the
momentum dependence of the mass operator) and $E-mass$ (characterizes the
explicit energy dependence of the mass operator), 
a third effective mass,
the "Lorentz mass" appears in the relativistic approach.
It results from different Lorentz transformation properties of
the scalar and vector potentials~\cite{JM.89,JM.90}.
However, what is usually termed "the relativistic effective mass" 
that characterizes an effective interaction, is a fourth mass:
the Dirac mass~\cite{JM.89}
\begin{equation}
m_D = m + S({\bm r})\; ,
\end{equation}
where $m$ is the nucleon mass and $S({\bm r})$ is the scalar 
nucleon self-energy. It has to be emphasized 
that the Dirac mass should not be identified
with the effective mass determined from non-relativistic shell
and optical model analyses of experimental data, i.e. with
the "non-relativistic-type effective mass". The Dirac mass 
is determined, on one hand by the binding energy 
at saturation density in nuclear matter (the effective 
single-nucleon potential is the sum of the attractive
scalar and repulsive vector nucleon self-energies), and 
on the other hand by the empirical spin-orbit splittings
in finite nuclei (the effective single-nucleon spin-orbit
potential is proportional to the difference between the
scalar and vector self energies). This is the reason why,
for virtually all mean-field relativistic effective
interactions, $0.55 m \leq m_D \leq 0.60 m$.

In contrast to the non-relativistic effective mass, the 
Dirac mass cannot be related to the ISGQR.
On the other hand, we would like to use the isoscalar quadrupole 
response to constrain the isoscalar properties of our
density-dependent effective interactions. This can be 
done in the following way. We first notice that a particular
ratio of isoscalar parameters $b_i$ and $c_i$ ($i\equiv \sigma,
\omega$) in Eq. (\ref{func}), characterizes the 
density of single-nucleon states around the Fermi surface. 
In Fig.~\ref{figF} we display the average energy gaps
between the last occupied and the first unoccupied major shells 
$<E_{\rm particles}> - <E_{\rm holes}>$ in $^{208}$Pb, where
\begin{equation}
<E> = {{\sum_{nlj} (2j+1) E_{nlj}}\over {\sum_{nlj}  (2j+1) }}\; ,
\end{equation}
and the sums run over occupied (unoccupied) states within a
major shell. The average gaps for proton and neutron states
are plotted as functions of the parameter $\delta$
\begin{equation}
\delta = \frac {b_\sigma/ c_\sigma}{b_\omega/c_\omega}\; ,
\label{delta}
\end{equation}
see Eq. (\ref{func}). Starting from DD-ME1, we have generated a set
of five effective interactions with $0.93 \leq \delta \leq 1.01$. 
For each interaction the remaining parameters were readjusted to 
reproduce our standard set of ground-state data for twelve 
spherical nuclei, as well as the nuclear matter equation of state.
The average gap between the last occupied and first unoccupied 
major shells, both for proton and neutron states, is approximately
linearly proportional to $\delta$. This parameter, therefore, plays 
the role of the inverse of the effective mass. As functions of $\delta$,
in Fig.~\ref{figG} we plot the corresponding centroid energy 
of the isoscalar quadrupole Hartree response in $^{208}$Pb
(upper panel), and the peak energies of the ISGQR 
obtained by the full RRPA calculation with the five density-dependent
interactions (lower panel). The calculated ISGQR excitation energies
are compared with the experimental value 
of $10.9\pm 0.3$ MeV \cite{Ber.79}
(shaded area). Both the centroids of the Hartree response and the 
ISGQR peak energies are linearly proportional to $\delta$ and the 
comparison with experimental data on ISGQR, therefore, places an 
additional constraint on the parameters that characterize the 
isoscalar channel of the effective interaction. For $\delta=0.93$, 
in Fig.~\ref{figH} we plot the RRPA isoscalar quadrupole strength
distribution in $^{208}$Pb (left panel), and for the 
ISGQR peak at 11.2 MeV the proton, neutron, and total isoscalar 
transition densities. The position of the calculated peak
should be compared with the empirical excitation energy
$10.9\pm 0.3$ MeV, and also the 
$0 \hbar\omega$ low-lying discrete $2^+$ state at 4.62 MeV is found
in good agreement with the experimental value of 4.07 MeV.


\section{\label{secIV}Summary and conclusions}

During the last decade the 
standard RMF models with nonlinear meson-exchange
effective interactions have been very successfully applied
in the description of a variety of nuclear structure phenomena.
In the last couple of years also the relativistic random-phase
approximation (RRPA), based on effective Lagrangians with
nonlinear meson self-interaction terms, has been used to
investigate properties of low-lying collective states and
of giant resonances. The use of nonlinear effective interactions,
however, presents not only a number of technical problems,
but also the predictive power of models based on these type
of interactions appears to be somewhat limited,
especially for isovector properties of exotic nuclei far from
$\beta$-stability. An interesting alternative are models 
with density-dependent meson-nucleon vertex functions. 
Even though these two classes of models are essentially based 
on the same microscopic structure, i.e. on density dependent 
interactions, the latter can be more directly related to the
underlying microscopic nuclear interactions. In a number of 
recent analyses it has been also shown that relativistic effective 
interactions, with explicit density dependence of the meson-nucleon
couplings, provide an 
improved description of asymmetric nuclear matter, neutron
matter and nuclei far from stability.

Among the new structure phenomena observed or predicted
in nuclei far from stability, one of the most interesting is the 
evolution of the isovector dipole response in nuclei with a
large neutron excess. 
The multipole response of nuclei with large neutron excess has
been the subject of many theoretical studies in recent years,
and some predictions have been confirmed by very recent
experimental data on low-lying electric dipole strength
in neutron rich nuclei~\cite{Lei.01,Zil.02}. There are, however,
many unknowns and this topic presents an interesting challenge
for modern theoretical advances. It is, therefore, important
to develop also a relativistic framework, based on effective
Lagrangians with density-dependent meson-nucleon couplings, 
in which the dynamics of exotic collective modes in nuclei 
far from stability can be investigated.

In this work we have derived the RRPA matrix equations in the 
small amplitude limit of the time-dependent relativistic mean-field
theory. The explicit density dependence of the meson-nucleon 
vertices introduces a number of rearrangement terms in the residual
two-body interaction. We have found that the rearrangement
contribution to the matrix elements of the RRPA equations is 
crucial for a quantitative comparison with experimental data
on giant resonances. In the present analysis we have performed
illustrative RRPA calculations of the isoscalar monopole,
isovector dipole and isoscalar quadrupole response of $^{208}$Pb.
The calculations are fully self-consistent: the single-particle
basis and the particle-hole couplings are generated from the same
effective Lagrangian, and the RRPA configuration space includes 
both the positive-energy particle-hole pairs, as well as pairs 
formed from hole states and negative-energy states in the 
Dirac sea. On one hand, we have tested our approach by comparing the
RRPA results for giant resonances with well known experimental data.
On the other hand we have also analyzed how the
RRPA results on multipole giant resonances can be used to 
constrain the parameters that characterize the isoscalar 
and isovector channel
of the density-dependent effective Lagrangians. Starting with
the recently introduced effective interaction DD-ME1~\cite{NVFR.02},
RRPA calculations have been performed for families of 
density-dependent interactions with a given characteristic 
(nuclear matter incompressibility, asymmetry energy etc.).

The analysis of the isoscalar monopole response has shown
that only the density-dependent interactions with
the nuclear matter compression modulus in the range 
$K_{\infty}\approx 260 - 270$ MeV, reproduce the experimental 
excitation energy of the isoscalar giant monopole resonance
in $^{208}$Pb. This confirms our previous results obtained with 
relativistic effective forces with non-linear meson 
self-interactions and points, once again, to the pronounced
difference  between the values of the
nuclear matter compression modulus predicted
by microscopic non-relativistic and relativistic
mean-field plus RPA calculations.
The RRPA results for the isovector dipole response constrain
the isovector channel of the effective interactions. By using
interactions with different values of the volume asymmetry
energy $a_4$, but which otherwise reproduce the same data
set of ground-state properties of spherical nuclei, we have
shown that the calculated IVGDR peak energy actually 
decreases by increasing the asymmetry energy at saturation.
The comparison with the experimental IVGDR 
excitation energy constrains the volume asymmetry to the
interval 34 MeV $\leq a_4 \leq$ 36 MeV. In the non-relativistic
framework the isoscalar quadrupole response can be related
to the effective mass of the mean-field interaction. 
The concept of effective mass in the relativistic mean-field
models is more complicated, and the quantity which is usually
termed as ``effective mass'' cannot be identified 
with the effective mass determined from non-relativistic shell
and optical model analyses of experimental data. Nevertheless,
we have shown that a comparison of RRPA results with the empirical
ISGQR and with the low-lying $0 \hbar\omega$  $2^+$ state,
places an additional constraint on the parameters which 
characterize the isoscalar channel of the density-dependent
effective interactions.

The RRPA with density-dependent meson-nucleon couplings
presents an important step in the relativistic description
of the nuclear many-body problem. In the present work we
did not attempt an analysis of the multipole response in exotic
nuclei far from $\beta$-stability. In order to do that,
pairing correlations must be included in the RRPA framework.
Work is in progress on the fully self-consistent relativistic
quasiparticle random-phase approximation (RQRPA), based
on effective Lagrangians with density-dependent meson-nucleon
couplings, and formulated in the relativistic Hartree-Bogoliubov
canonical single-particle basis.

\bigskip
\bigskip
\leftline{\bf ACKNOWLEDGMENTS}

This work has been supported in part by the
Bundesministerium f\"ur Bildung und Forschung under
project 06 TM 979, and by the Gesellschaft f\" ur
Schwerionenforschung (GSI) Darmstadt.
T. N. acknowledges the support from the Alexander von
Humboldt - Stiftung.

==========================================================================

\newpage
\begin {table}[]
\begin {center}
\caption {The effective interaction DD-ME1. See Eqs. (\protect\ref{coupl})--
(\protect\ref{grho}) for the definition of the coupling parameters.}
\begin {tabular}{cc}
          &          DD-ME1      \\ \hline
\hline
$m_{\sigma}$             & {549.5255 }   \\
$m_{\omega}$             & {783.0000 }   \\
$m_{\rho}  $             & {763.0000 }   \\
$g_{\sigma}(\rho_{sat})$ & { 10.4434 }   \\
$g_{\omega}(\rho_{sat})$ & { 12.8939 }    \\
$g_{\rho}(\rho_{sat})  $ & {  3.8053 }   \\
$a_{\sigma}$             & {  1.3854 } \\
$b_{\sigma}$             & {  0.9781 }   \\
$c_{\sigma}$             & {  1.5342 } \\
$d_{\sigma}$             & {  0.4661 } \\
$a_{\omega}$             & {  1.3879 } \\
$b_{\omega}$             & {  0.8525 } \\
$c_{\omega}$             & {  1.3566 } \\
$d_{\omega}$             & {  0.4957 } \\
$a_{\rho}$               & {  0.5008 } 

\end {tabular}
\label{tab1}
\end{center}
\end{table}
\newpage
\begin{figure}
\caption{\label{figA} Density-dependent RRPA peak energies of the
ISGMR in $^{208}$Pb as a function of the nuclear matter compressibility
$K_\infty$.
The calculated peaks are shown in comparison with the experimental 
excitation energy of the monopole resonance: $E = 14.1\pm 0.3$ MeV
\protect\cite{YCL.99}.}
\end{figure}
\begin{figure}
\caption{\label{figB} The isoscalar monopole
strength distribution (left panel) and
transition densities (right panel) in $^{208}$Pb, calculated with 
a density-dependent effective interaction with $K_\infty = 270$ MeV.
The proton, neutron and total isoscalar transition 
densities correspond to the peak at $E = 14.1$ MeV.}
\end{figure}
\begin{figure}
\caption{\label{figC} The IVGDR excitation energy of $^{208}$Pb (upper
left panel), and the parameter $p_0$ of the linear density dependence 
of the nuclear matter asymmetry energy, as 
functions of the volume asymmetry $a_4$. The shaded area denotes the 
experimental IVGD resonance energy $13.3\pm 0.1$ MeV. In the right panel
the asymmetry energy curves, as functions of the baryon density, are 
plotted for different values of the volume asymmetry $a_4$.}
\end{figure}
\begin{figure}
\caption{\label{figD} The isovector dipole
strength distribution (left panel) and
transition densities (right panel) in $^{208}$Pb, calculated with 
a density-dependent effective interaction with $a_4 = 35$ MeV.
The proton, neutron and total isovector transition 
densities correspond to the peak at $E = 13.3$ MeV.}
\end{figure}
\begin{figure}
\caption{\label{figE} The IVGDR excitation energy of $^{208}$Pb (upper
left panel), and the parameter $p_0$ of the linear density dependence 
of the nuclear matter asymmetry energy, as 
functions of the volume asymmetry $a_4$. The shaded area denotes the 
experimental IVGD resonance energy $13.3\pm 0.1$ MeV. In the right panel
the asymmetry energy curves, as functions of the baryon density, are 
plotted for different values of the volume asymmetry $a_4$.}
\end{figure}
\begin{figure}
\caption{\label{figF} Average energy gap between the last occupied 
and first unoccupied major shells in $^{208}$Pb, as function 
of the isoscalar parameter $\delta$ (\protect\ref{delta}).
The average gaps of neutron states are denoted by dots, and
those of proton states by squares.}
\end{figure}
\begin{figure}
\caption{\label{figG} Centroid energies of the isoscalar quadrupole
Hartree response (upper panel), and the ISGQR peak energies 
calculated in RRPA (lower panel), 
for five different density-dependent interactions
characterized by the parameter $\delta$ (\protect\ref{delta}).
As in the previous examples, the calculation is performed 
for $^{208}$Pb. The shaded area denotes the empirical 
ISGQR excitation energy in $^{208}$Pb: $10.9 \pm 0.3$ MeV
\protect\cite{Ber.79}.}
\end{figure}
\begin{figure}
\caption{\label{figH} The isoscalar quadrupole strength distribution
(left panel) and transition densities (right panel) in $^{208}$Pb,
calculated for $\delta=0.93$ (see text for description). The vertical
bar denotes the $0 \hbar\omega$  $2^+$ discrete state. The 
proton, neutron, and total isoscalar transition densities correspond
to the ISGQR peak at $E=11.2$ MeV excitation energy.}
\end{figure}
\end{document}